\documentclass[10pt]{iopart}
\usepackage{graphicx}
\usepackage{array}
\usepackage{url}
\usepackage{hyperref}
\usepackage{bm}
\usepackage{iopams}

\begin{document} \title{Reply to comment on ``Improvements for drift-diffusion
plasma fluid models with explicit time integration''}

\author{Jannis Teunissen$^{1}$}

\address{$^1$Centrum Wiskunde \& Informatica (CWI), P.O. Box 94079, 1090 GB
Amsterdam, The Netherlands}

\ead{jannis@teunissen.net}

\begin{abstract}
  This is a reply to the comment of Jiayong Zou on the paper ``Improvements for
  drift-diffusion plasma fluid models with explicit time integration''. The
  criticism in the comment, namely that the current-limited approach is
  inconsistent with the underlying partial different equations, seems to be
  invalid. However, this criticism raises an interesting question about the
  behavior of the current-limited scheme for a given time step, which is
  discussed in this reply.
\end{abstract}

\maketitle \ioptwocol

\section{Reply to comment}

The comment by J.~Zou~\cite{Zou_2020} raises concerns about the consistency of the
current-limited scheme described in \cite{Teunissen_2020}. A numerical scheme is
\emph{consistent} when numerical solutions converge to the true solution of the
underlying partial differential equation (PDE) when the mesh spacing $\Delta x$ and
time step $\Delta t$ go to zero. A similar definition is also given in the
comment.

The main criticism of the comment is that the scheme proposed in
\cite{Teunissen_2020} is not consistent. However, there seems to be a flaw in
the reasoning: the case $\Delta x \to 0$ is considered without taking into
account $\Delta t \to 0$. According to equation (18) of \cite{Teunissen_2020},
the electron flux is only limited when its amplitude exceeds
\begin{equation}
  \label{eq:fmax} f_\mathrm{max} = \varepsilon_0 E^*/(e \Delta t),
\end{equation}
where $\varepsilon_0$ is the permittivity of vacuum, $e$ the elementary charge
and $E^*$ the amplitude of the electric field with a correction for diffusive
terms, see equation (19) of \cite{Teunissen_2020}. For $\Delta t \to 0$, we have
$f_\mathrm{max} \to \infty$, so the electron flux will not be limited or
otherwise modified. The proposed scheme is therefore consistent with the
underlying PDE.

\section{Behavior of current-limited scheme}

Regardless of whether the proposed scheme is consistent, the comment does raise
an interesting question: how does the current-limited scheme behave for a given
time step $\Delta t$? To analyze the scheme, let us consider a simple homogeneous plasma, with the following assumptions:
\begin{itemize}
  \item The electron density $n_e$ is homogeneous in space and time
  \item The electron mobility $\mu_e$ is constant
  \item The conductivity only comes from electrons
  \item There is no space charge, and the electric field $E$ in the plasma is
  homogeneous
\end{itemize}
Physically, the electron flux will then be given by
\begin{equation}
  \label{eq:f}
  f = -n_e \mu_e E,
\end{equation}
and the dielectric relaxation time is given by
\begin{equation}
  \label{eq:tau}
  \tau = \frac{\varepsilon_0}{e \mu_e n_e}.
\end{equation}
A standard numerical discretization will give the same flux as equation
(\ref{eq:f}). A semi-implicit
scheme~\cite{Ventzek_1994,Lapenta_1995a} will also reproduce this
flux, since the `correction' term will be zero, see equation (15) in
\cite{Teunissen_2020}. Let us now consider the current-limited approach, for
which $f_\mathrm{max}$ is here given by
\begin{equation}
  \label{eq:fmax} f_\mathrm{max} = \varepsilon_0 |E|/(e \Delta t).
\end{equation}
With a time step $\Delta t \leq \tau$ we will have $|f| \leq f_\mathrm{max}$, so
no limiting will take place and the correct flux is obtained. For
$\Delta t > \tau$, the current-limited scheme will reduce $|f|$ so that it does
not exceed $f_\mathrm{max}$, effectively reducing the conductivity of the plasma
to prevent instabilities. Depending on the applied boundary condition, this can lead to two effects:
\begin{itemize}
  \item When a constant voltage is applied over the plasma, the current through the
  plasma will reduce by a factor $\tau / \Delta t$.
  \item When a constant current is imposed through the plasma, the electric
  field in the plasma will increase by a factor $\Delta t / \tau$.
\end{itemize}
These unphysical effects are drawbacks of the current-limited scheme, but there seems to be no way to overcome them without making the scheme at least partially implicit.

The example above highlights the drawbacks of current-limited scheme because it
contains a `short-circuit' and therefore a potentially large current. The scheme
is more suitable for discharges that do not carry large currents, like the
examples shown in \cite{Teunissen_2020}, in which the electric field is screened
inside the plasma. Increasing such a screened electric field by a factor
$\Delta t / \tau$, as described above, will not significantly affect the
obtained solution.

The scheme can sometimes also be used when a current flows through the plasma.
If a small region has a much higher conductivity than the rest of the plasma,
the conductivity inside the small region will be reduced, and the local electric
field will increase correspondingly. However, the total conductivity of the
plasma will remain almost the same.

In summary, the current-limited scheme artificially reduces the conductivity of
the plasma when $\Delta t > \tau$. How severely this will impact solutions
depends on the type of system that is simulated and on the ratio
$\Delta t/\tau$. The examples in~\cite{Teunissen_2020} demonstrate that in some
regimes, the errors introduced by the current-limited scheme are significantly
smaller than those resulting from a first-order semi-implicit scheme.

\section*{References}
\bibliographystyle{elsarticle-num}
\bibliography{refs}

\end{document}